\documentclass{emulateapj}
\usepackage{apjfonts}

\begin{document}

\title{Polarization signature of gamma-ray bursts from fragmented 
fireballs}
\shorttitle{GRB polarization}

\author{Davide Lazzati\altaffilmark{1} and 
Mitchell C. Begelman\altaffilmark{2,3}}
\email{davide\_lazzati@ncsu.edu}
\shortauthors{Lazzati \& Begelman}

\altaffiltext{1}{Department of Physics, NC State University, 2401
Stinson Drive, Raleigh, NC 27695-8202} 
\altaffiltext{2}{Jila, University of Colorado, 440 UCB, Boulder, CO
80309-0440}
\altaffiltext{3}{University of Colorado, Department of Astrophysical and 
Planetary Sciences, 389 UCB, Boulder, CO 80309-0389}

\begin{abstract}
We study the polarization properties of the prompt emission of gamma-ray
bursts produced by fragmented fireballs. Such fireballs, known in the
literature under various names, are made by the superposition of many
individual blobs, each of which produces a spike in the light curve. The
differences between pulses are due both to the intrinsic diversity in
the blobs' properties and to their orietation with respect to the line
of sight. We show that the peak flux and the polarization of each pulse
are connected through the orientation of the blob that produces the
pulse, while the position angle fluctuates randomly from one pulse to
the next and is constant within an individual pulse. The most polarized
pulses are those with approximately one tenth the peak flux of the
brightest pulse. These conclusions do not depend on the assumed
radiation mechanism nor on the energy and Lorentz factor of the blobs.
We compare the prediction of this model to a simulated set of
observations, showing that a limited sample of GRBs with time-resolved
polarization measurements would provide a crucial test for this model.
We finally show that a hint of the predicted correlation may have been
already observed.
\end{abstract}


\section{Introduction}

Despite many years of observations, the origin of the prompt emission of
gamma-ray bursts (GRBs) is still clouded in mystery. GRB prompt
radiation is known to originate within a collimated relativistic outflow
with Lorentz factor of several hundreds (Rhoads 1999; Lithwick \& Sari
2001 and references therein). However, the geometry of the outflow,
whether it is matter or Poynting flux dominated, the dissipation
processes, and the radiation mechanism involved are still matters of
open debate.

The amount and the temporal evolution of linear polarization of the
prompt GRB emission has been hailed as a key observation to solve, at
least, some of these outstanding issues (Eichler \& Levinson 2003;
Granot 2003; Lyutikov et al. 2003;  Nakar et al. 2003; Waxman 2003;
Lazzati et al. 2004a; Lazzati 2006; Dado et al. 2007). On the
observational side, the measurement of linear polarization at several
hundred keV has been challenging. An initial report of very high linear
polarization in the prompt emission of GRB~021206 ($\Pi=80\pm20\%$;
Coburn \& Boggs 2003) was subsequently shown to be very uncertain, at
best (Rutledge \& Fox 2004; Wiggers et al. 2004). Subsequent claims of
moderate to high levels of linear polarization in the prompt emission of
other GRBs have been plagued by low significance and  systematic
uncertainties: $\Pi>35\%$ in GRB 930131 and $\Pi>55\%$ in GRB 960924
(Willis et al. 2005); $\Pi=98\pm33\%$ in GRB 041219a (Kalemci et al.
2007; McGlynn et al. 2007).

Most of the models proposed to explain the high linear polarization
possibly observed predict a constant position angle, since the
polarization is expected to lie in the plain that contains both the axis
of the jet and the line of sight to the observer. However, a recent
observation by Gotz et al. (2009) of GRB 041219a suggests that the
polarization angle changes substantially within the burst. Prompted by
this observation, we explore in this letter the polarization arising
from fragmented fireballs, i.e., fireballs that are made by the
superposition of smaller jets or relativistic blobs of material. Within
this framework, each pulse of a light curve is produced by a single
blob, and the pulse properties depend on the blob intrinsic properties
(energy and Lorentz factor) as well as on the orientation between the
velocity vector of the blob and the line of sight to the observer.
Fragmented fireballs have been proposed in the past under various names
as a way to explain GRB prompt emission: shotguns (Heinz \& Begelman
1999), cannonballs (Dado \& Dar 2009 and references therein), sub-jets
(Yamazaki et al. 2006). Additional models that predict a polarization
behavior analogous to fragmented fireballs are external shocks on a
clumpy medium (Dermer et al. 1999) and precessing jets (Blackman et al.
1996; Portegies Zwart et al. 1999).

This paper is organized as follows: in \S~2 we compute the polarization
arising from a fragmented fireball, in \S~3 we compare the predictions
to observations and in \S~4 we discuss our results and compare them to
the predictions of alternative models.

\section{Polarization from a single fragment}

Consider a blob with total energy $E$ moving with Lorentz factor
$\Gamma$ with an angle $\theta_o$ between the velocity vector and the
line of sight. Both the radiation flux received at earth and the
polarization fraction depend on the viewing angle $\theta_o$.

If the blob releases a certain fraction of its energy into radiation,
the photon pulse appears brightest for an observer at $\theta_o=0$.
We call the flux of photons from the on-axis blob in photons per
second per square centimeter $\Phi_{\max}$. If the same blob moves with an
angle $\theta_o>0$, its photon flux scales as
\begin{equation}
\Phi(\theta_o) = \Phi_{\max} \frac{\delta(\theta)^3}{\delta(0)^3} \simeq
\frac{\Phi_{\max}\delta(\theta)^3}{8\Gamma^3}
\label{eq:iimax}
\end{equation}
where $\delta(\theta)=[\Gamma(1-\beta\cos\theta)]^{-1}$ is the Doppler
factor.

Consider now radiation produced either by synchrotron emission in a
shock-compressed magnetic field (Ghisellini \& Lazzati 1999; Sari 1999;
Granot 2003) or by bulk inverse Compton scattering (Begelman \& Sikora
1987; Shaviv \& Dar 1995, Lazzati et al. 2004). In either case, the
radiation has a polarization 
\begin{equation}
\Pi(\theta^\prime)=\Pi_{\max} \frac{1-\cos^2\theta^\prime}
{1+\cos^2\theta^\prime}
\label{eq:ppmax}
\end{equation}
where $\theta^\prime$ is the angle, in the comoving frame, between the
velocity vector and the direction of the photon velocity and
$\Pi_{\max}$ is the maximum polarization attainable by the model:
$100\%$ for inverse Compton and $\sim 70\%$ for synchrotron.

\begin{figure}
\plotone{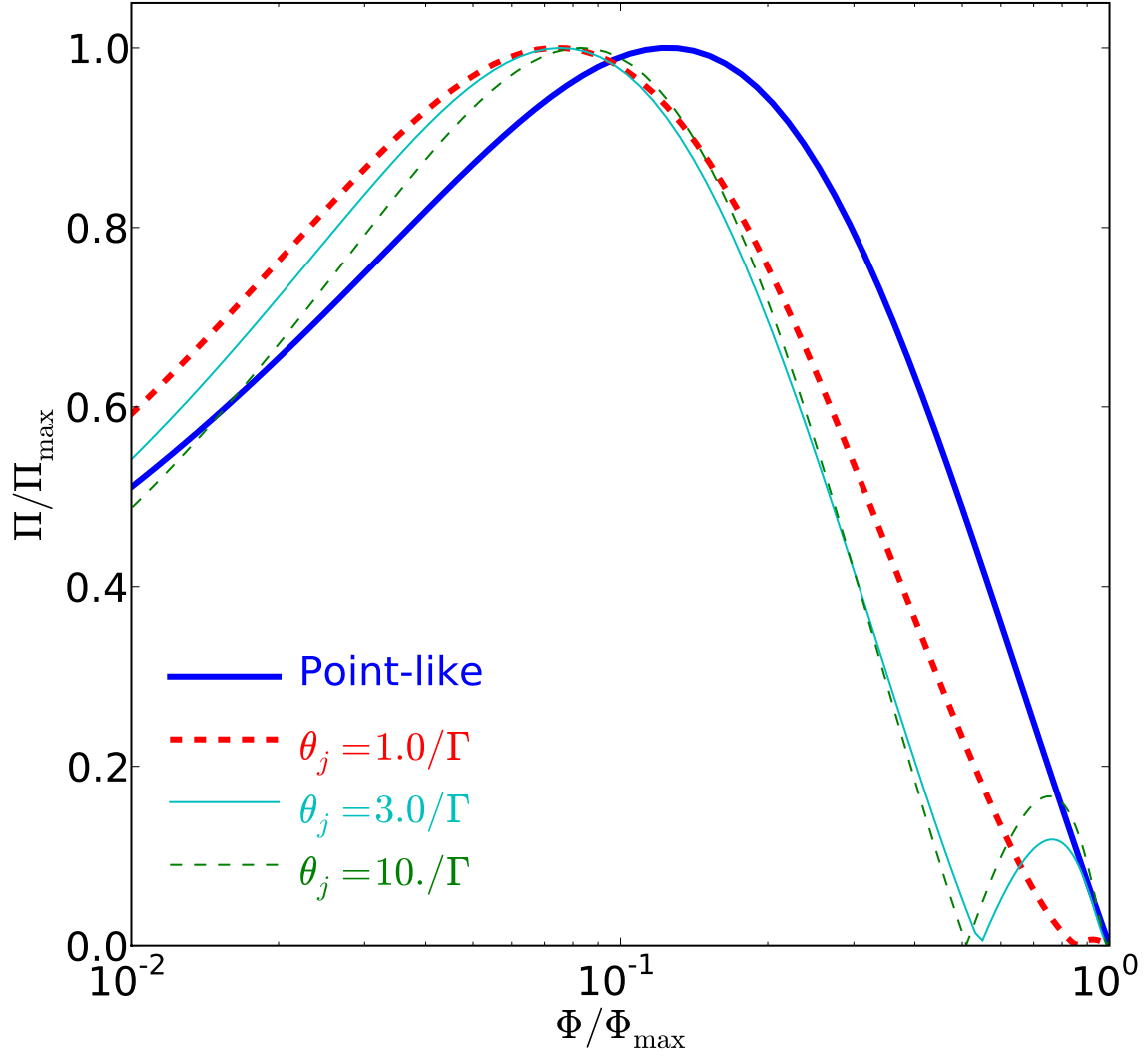}
\caption{{Polarization versus photon peak flux for a fragmented
fireball. The thick solid line (blue) shows the point-like approximation,
for which $\theta_j<<1/\Gamma$. The thick dashed line (red) shows a
causally connected fireball ($\theta_j=1/\Gamma$), the thin solid line (cyan)
shows a fireball with $\theta_j=3/\Gamma$, and the thin dashed line
(green) shows a fireball with $\theta_j=10/\Gamma$).}
\label{fig:1}}
\end{figure}

Due to the relativistic aberration, the comoving angles map into
observer's angles as 
\begin{equation} 
\cos\theta^\prime =
\frac{\cos\theta-\beta}{1-\beta\cos\theta} 
\label{eq:aberr}
\end{equation} 
where $\beta$ is the velocity of the fireball in units of the speed of
light. Equations~\ref{eq:iimax}, \ref{eq:ppmax}, and~\ref{eq:aberr} are
an implicit system of equations that allow us to plot the fraction
$\Pi/\Pi{\max}$ as a function of $\Phi/\Phi_{\max}$, independently of
the fragment energy, radiation mechanism, and Lorentz factor. The result
is shown with a thick solid line in Fig.~\ref{fig:1}. The brightest
configuration has no polarization since the fragment points directly to
the observer and everything is symmetric. Polarization increases with
$\theta_o$ until the maximum is reached when $\theta_o\sim1/\Gamma$
(which corresponds to $\theta^\prime=\pi/2$ in the comoving frame) and
$\Phi\sim \Phi_{\max}/8$.

The analytic result shown above is rigorously valid for a point-like
fragment, for which the opening angle $\theta_j\ll1/\Gamma$. In
realistic cases, we expect $\theta_j\sim1/\Gamma$ or larger, since a hot
fragment would expand into the cone of causal connection. We compute the
curves of $\Pi/\Pi_{\max}$ versus $\Phi/\Phi_{\max}$ for realistic
fragments following Lazzati et al. (2004) and we show the results in
Fig.~\ref{fig:1}. The behavior is analogous, the main difference being a
region of low polarization for $\Phi_{\max}/2<\Phi<\Phi_{\max}$. This is
the range of angles for which the line of sight is still inside the
fragment, but is not aligned with the fragment axis.

The electric field vector, which defines the position angle of the
polarization, lies in the plane that contains both the line of sight to
the observer and the velocity vector of the blob. As a consequence, the
position angle is constant for an individual pulse. However, since
different pulses are due to different blobs that do not share the same
velocity vector, the polarization angle should fluctuate randomply from
one pulse to the next. Such behavior is not possible in polarization
models for non-fragmented fireballs, since the velocity vector in that
case maintains the same direction during the whole prompt emission
phase.

\section{Observational considerations}

The results discussed above allow us to predict the polarization of the
pulse from a fragment as a function of the reduction of the  peak photon
flux with respect to the fragment pointing at the observer. If a GRB
fireball were made of identical fragments (identical energy, opening
angle, and Lorentz factor) and if at least one of the fragments would
always be pointing directly to the observer, polarization measurements
from a GRB would lie along one of the lines plotted in Fig.~\ref{fig:1}.

In reality, fragmented fireball models predict fragments with different
energies and Lorentz factors and different opening angles. In addition,
the brightest observed pulse may not come from a perfectly aligned
fragment and therefore the value of $\Phi_{\max}$ may be misidentified.
In order to understand the impact of this diversity, we performed a
Monte Carlo simulation in which we considered a sample of 30 bursts with
the following characteristics. All bursts had 29 fragments with energy
randomly distributed between 0.1 and 1 in arbitrary units. The fragments
had Lorentz factor randomly distributed between 100 and 400 and opening
angle $\theta_j=1/250=1/\langle\Gamma\rangle$. All the fragments were
ejected within a collimated outflow with opening angle 0.02 radians. In
such a configuration, an observer along the axis of the outflow sees on
average 4 pulses within 10\% of the peak photon flux. 

The result of the simulation are shown in Fig~\ref{fig:2}. The yellow
dots show $\Pi/\Pi_{max}$ of each individual pulse versus its
$\Phi/\Phi_{\max}$. Note that $\Phi_{\max}$ and $\Pi_{\max}$ are calculated
for the GRB to which the pulse belongs and not from the theoretical
maxima. The yellow dots show a big dispersion around the theoretical
curve for a fragment with $\theta_j=1/\Gamma$ (shown with a red solid
line) and have very little predictive power. However, the average
behavior of the sample is strongly indicative of the underlying
mechanism that produces the polarization. Blue symbols show the average
value of $\Pi/\Pi_{\max}$ in 10 bins of $\Phi/\Phi_{\max}$. Even though
the modulation of the polarization is not as deep as for the point-like
blob, a trend of increasing polarization for decreasing peak flux is
clearly observable and highly statistically significant (error bars show
the standard error of the sample in each bin). Finally, the green line
shows the average results that would be obtained with a much larger GRB
sample (30000) for which the uncertainty is negligible. Even a large
sample does not reproduce the ideal case, mainly due to the low
probability of having a blob pointing directly to the observer with only
4 bright pulses in the lightcurve.

The simulation of Fig.~\ref{fig:2} shows that even a moderate number of
GRBs with reliable time-resolved polarization measurements would
represent a crucial test for this model. We should emphasize here that
the simulation is made conservatively, since in the real case selection
effects would tend to privilege bursts with at least one very bright
peak. In such a case, a better evaluation of $\Phi_{\max}$ would be
possible. In addition, any burst with more than 4 pulses, would allow
for a better determination of $\Phi_{\max}$.

\begin{figure}
\plotone{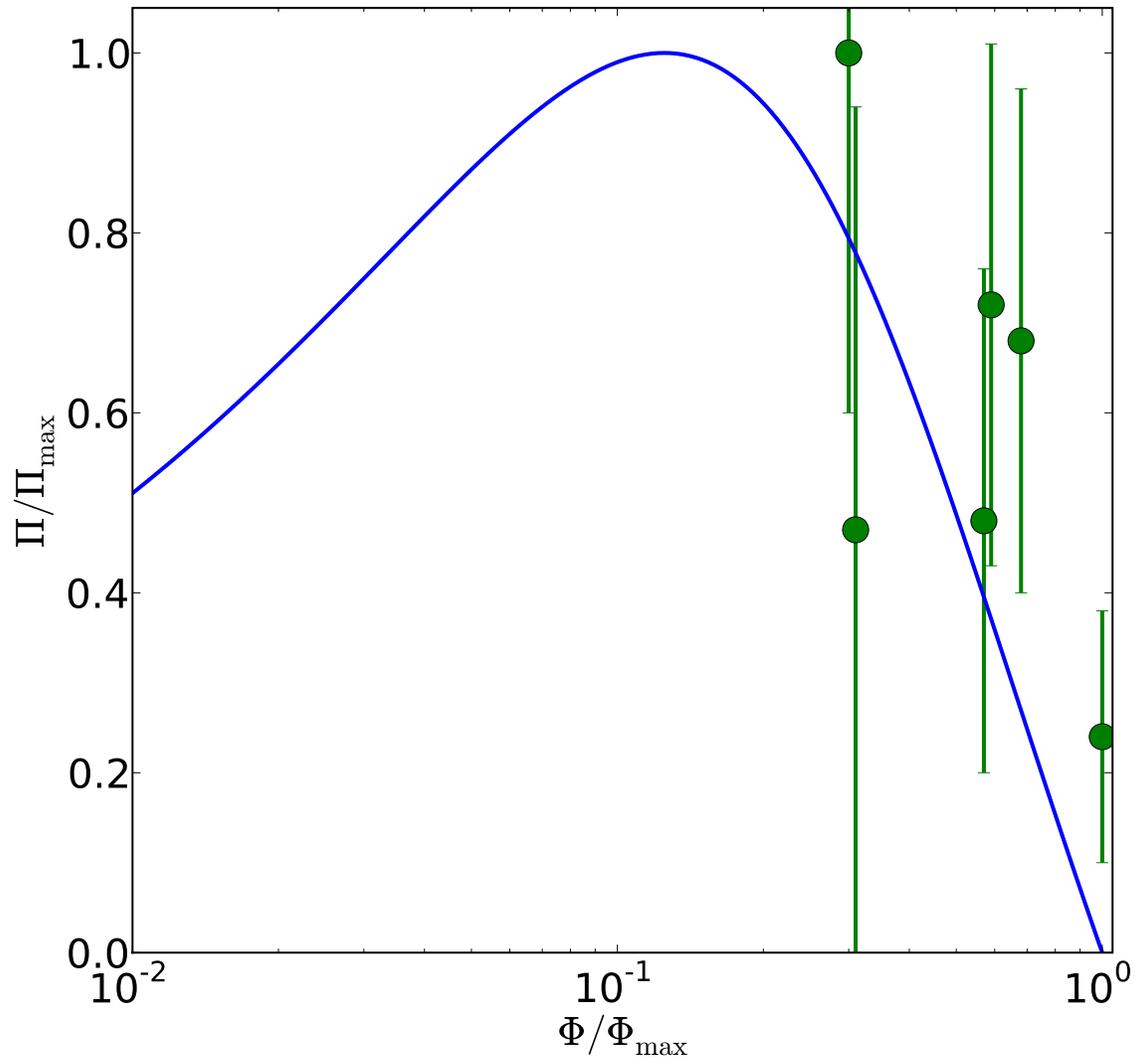}
\caption{{Polarized fraction of GRB 041219a from G\"otz et al. (2009)
overlaid on the point-like prediction for the fragmented fireball model.}
\label{fig:3}}
\end{figure}

In Fig.~\ref{fig:3} we show the polarization results of G\"otz et al.
(2009) overlaid on the model predictions. Even though the error bars are
big and strong conclusions cannot be drawn, the observations suggest
that the bright peaks are less polarized than the dim ones. However, a
systematic effect may produce this result, since dim peaks have less
statistics and would produce more uncertain polarization measurements.
More robust measurements are necessary to confirm the suggestive
Fig.~\ref{fig:3}.

\section{Discussion}

We have shown that a fragmented fireball can easily produce highly
polarized GRBs with a position angle that fluctuates randomly among
individual pulses. In this model, the brightest pulses should be the
least polarized, with the maximum polarization observed for pulses with
peak photon flux of about $10\%$ of the peak photon flux of the GRB.
These characteristics, possibly observed by G\"otz et al. (2009) in GRB
041219a, are not reproduced by alternative scenarios:

\begin{itemize}

\item {\bf Magnetic Domains.} Gruzinov \& Waxman (1998) proposed a model
in which the magnetic field behind the shock re-organizes into magnetic
domains. Within each domain, the field is uniform and gives rise to
maximally polarized radiation. Even though the model can explain a
randomly fluctuating angle, the maximum polarization expected is of
$\Pi\le10\%$, since approximately 100 domains are simultaneously visible
to the observer, even if the re-organization proceeds at the speed of
light. In addition, this model is contradicted by afterglow observations
that show a position angle that varies within a small interval (Lazzati
et al. 2004b; Greiner 2004).

\item $\mathbf{1/\Gamma}$ {\bf effects} A number of models predict high
polarization in the prompt GRB emission as a consequence of a particular
viewing geometry (Waxman 2003; Granot 2003; Lazzati et al. 2004a). In
such models, the radiation mechanism (either synchrotron or bulk inverse
Compton) produces highly polarized radiation in the comoving frame in
the direction perpendicular to the direction of motion. In the observer
frame, due to relativistic aberration, the highly polarized radiation is
observed in a direction that lies at an angle $\theta=1/\Gamma$ with
respect to the velocity vector. If the opening angle of the fireball is
small ($\theta_j\le1/\Gamma$), such an observer could detect
polarization up to 100\%. However, the polarization position angle lies
in the plane that contains the line of sight and the velocity vector,
and therefore it can not change during the burst.

\item {\bf Toroidal magnetic field}. If the magnetic field component is
dominant in the energy budget of the jet, the GRB radiation is produced
by synchrotron emission in a toroidal magnetic field and the
polarization can be up to ~50\%, regardless of the particular
configuration of the fireball with respect to the line of sight
(Lyutikov et al. 2003). Again, even though this model predicts high
linear polarization in most GRBs, it cannot account for position angle
variations within a single burst, since the polarization is always
directed towards the jet axis.

\end{itemize} 

\begin{figure}
\plotone{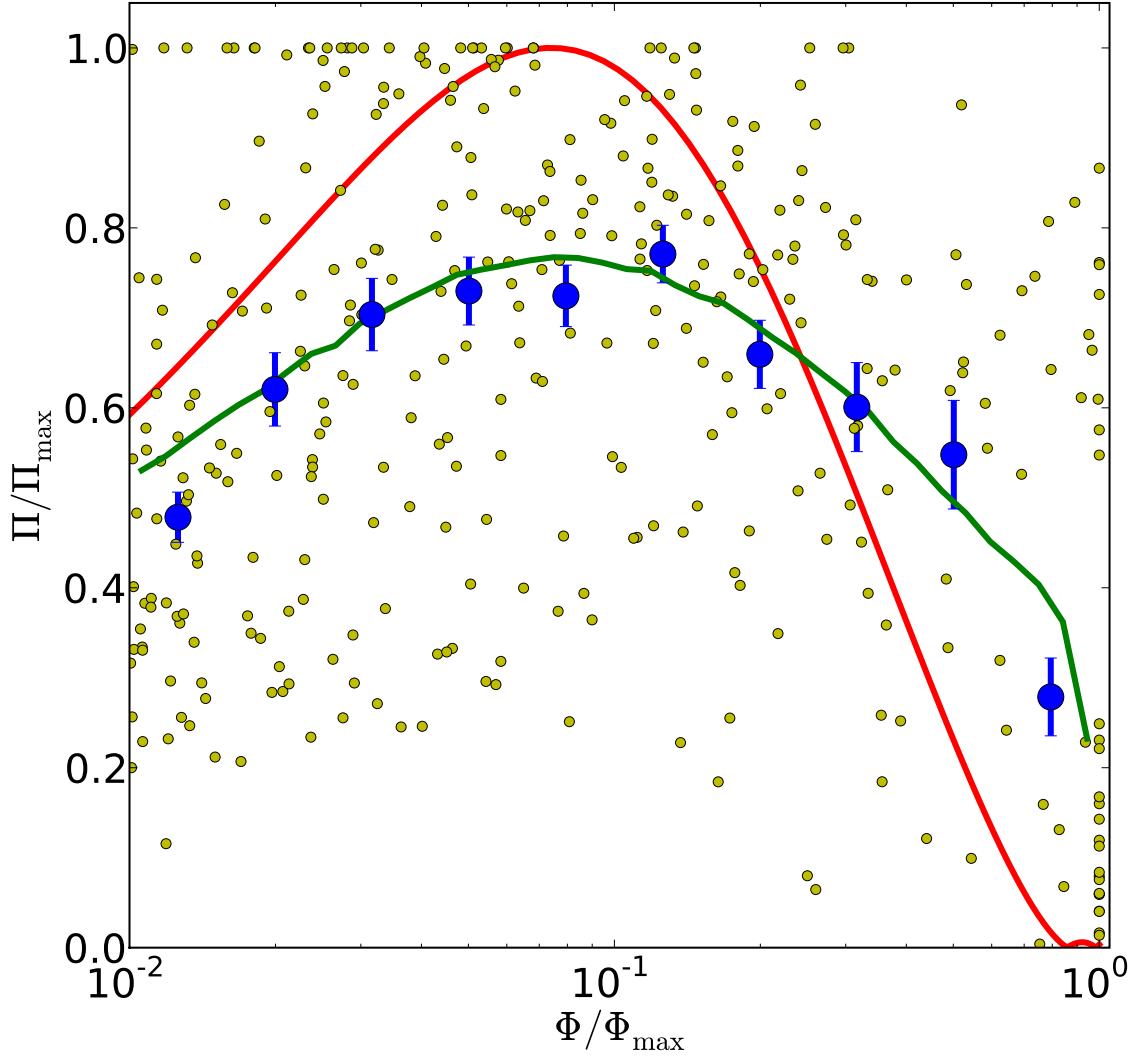}
\caption{{Simulation of the observation of a sample of 30 GRBs (yellow
circles). Each of them has 5 peaks with peak flux within 10\% of the
maximum. The energy of each fragment is randomly distributed between 0.1
and 1 in arbitrary units, the Lorentz factor is randomly distributed
between 100 and 400 and each fragment has an opening angle
$\theta_j=1/250=1/\langle\Gamma\rangle$. The red solid line shows the
polarization versus intensity for the point-like approximation. The blue
symbols with error bars show the average polarization versus intensity
in 10 bins of intensity, while the green solid line shows the result for
a very large sample of GRBs.}
\label{fig:2}}
\end{figure}

For these reasons, should high linear polarization with a varying
position angle be confirmed by future observations, it would be a strong
indication that the fireball producing GRBs is fragmented or, at least,
that the energy release is not uniform within the jet opening angle.
This models comes with the prediction that the brightest pulses are
least polarized, since they are directed straight to the observer.
Pulses with approximately 10\% of the peak luminosity should be
maximally polarized, since they are directed at an angle $1/\Gamma$ with
respect to the line of sight. As shown in Fig.~\ref{fig:2}, the
prediction will be easily tested when a reasonably large number of GRBs
will have observations of linear polarization of their prompt emission.

\acknowledgements

We thank Gabriele Ghisellini for useful discussions. This work was in
part supported by NASA ATP grant NNG06GI06G (DL \& MB) and by Swift GI
program NNX08BA92G (DL).

\newpage

\newpage

\newpage

\end{document}